\documentstyle[aps,prbbib,twocolumn,epsf]{revtex}

\begin{document}
\draft
\title{Nonlinear Quantum Capacitance}

\author{Baigeng Wang$^1$, Xuean Zhao$^1$, Jian Wang $^1$ and Hong Guo$^2$}
\address{1. Department of Physics, The University of Hong Kong, 
Pokfulam Road, Hong Kong, China\\
2. Center for the Physics of Materials and Department 
of Physics, McGill University, Montreal, PQ, Canada H3A 2T8\\}
\maketitle

\begin{abstract}
{\bf 
We analyze the nonlinear voltage dependence of elelctrochemical capacitance
for nano-scale conductors. This voltage dependence is due to finite
density of states of the conductors. We derive an exact expression 
for the electrochemical capacitance-voltage curve for a parallel plate
system. The result suggests a {\it quantum scanning capacitance microscopy} 
at the nano-scale: by inverting the capacitance-voltage expression one is 
able to deduce the local spectral function of the nano-scale conductor.
}
\end{abstract}


It has been well known that density of states affects the capacitance\cite{smith} 
of a system. In this work we investigate a nonlinear bias voltage dependence 
of the electrochemical capacitance of a model capacitor at the nano-scale 
for which the density of states (DOS) also plays the crucial role. 
The physical origin of this bias dependence which we have examined, is {\it not} 
because of depletion of charges on a capacitor plate which has been understood 
in semiconductor research, but because of the finite DOS of the plates.
We show that the accumulated charge on a conductor has a nonlinear 
bias dependence due to DOS effects, it is this nonlinear charge which leads 
to the nonlinear capacitance. For conductors at nano-scale, it is 
known\cite{but1} that the very small DOS plays an important role for the 
behavior of capacitance. However the DOS induced nonlinear capacitance of 
nano-conductors has never been studied\cite{buttiker5} before and our 
investigation suggests a very interesting application which is the quantum 
scanning capacitance microscopy.

Central to the problem is the determination of charge $\rho_\alpha$
accumulated on a conductor labeled by $\alpha$. In general
$\rho_\alpha$ is a nonlinear function of bias voltages $\{V_\beta\}$,
\begin{eqnarray}
\rho_\alpha &=& \sum_\beta C_{\alpha \beta} V_\beta + \frac{1}{2} \sum_{\beta
\gamma} C_{\alpha \beta \gamma} V_\beta V_\gamma + ... \nonumber \\
&\equiv & \sum_\beta C_{\alpha \beta}(\{V_\gamma\}) V_\beta
\label{cap}
\end{eqnarray}
where $C_{\alpha \beta}=\partial_{V_\beta} \rho_{\alpha}$ is the usual 
electrochemical capacitance\cite{but1,wang}, $C_{\alpha \beta \gamma}=
\partial_{V_\beta} \partial_{V_\gamma} \rho_{\alpha}$\cite{ma} {\it etc} 
are the nonlinear electrochemical capacitance coefficients, and 
$C_{\alpha \beta}(\{V_\alpha\})$ is the general voltage dependent nonlinear 
capacitance. So far investigations on quantum correction to 
capacitance\cite{smith,but1,wang} have only considered the {\it linear}
term, but in this work we focus on a general nonlinear expression for
$C_{\alpha \beta}(\{V_\gamma\})$. 

To be specific we consider a model parallel plate capacitor connected to
electron reservoirs by perfect leads\cite{but1}. We adopt the dynamic point
of view\cite{but6} to calculate the electrochemical capacitance, hence a
finite bias applied at the reservoirs injects a charge density into the
capacitor plates which, through interaction, induces a local response. The
total net charge (the sum of injected and induced charge) 
at a plate $\alpha$, $\rho_\alpha$, is thus established at equilibrium.
We calculate $\rho_\alpha$ by extending the standard nonequilibrium Green's
function (NEGF) technique. To save space we refer
interested readers to literature for standard technical details, 
here we will only discuss the most important extension to the standard 
approach.

The quantum scattering (injection of electrons) 
is determined by the retarded and advanced Green's 
function $G^{r,a}(E,U)$: note that we have explicitly included the 
electro-static potential build-up inside our capacitor, $U=U({\bf r})$, 
into these Green's functions. In the Hartree approximation\cite{datta1},
\begin{equation}
G^{r,a}(E,U) = \frac{1}{E-H-qU-\Sigma^{r,a}}
\label{gr}
\end{equation}
where $H$ is the Hamiltonian for our nano-scale conductors written in the
familiar second quantized form\cite{wang2}, $\Sigma^{r,a}$ is the self-energy 
describing the coupling between the conductors to the leads which 
is calculated in standard fashion\cite{wingreen,datta}. 
At Hartree level we determine $U({\bf r})$ by the self-consistent
Poisson equation
\begin{equation}
\nabla^2 U = 4\pi i q\int (dE/2\pi)G^<(E,U)
\label{poisson}
\end{equation}
where the right hand side is just the total net charge distribution in our 
conductors. Within Hartree approximation,
\begin{equation}
G^<(E,U) = G^r \sum_{\beta} i \Gamma_{\beta}(E-qV_{\beta})
f(E-qV_{\beta}) G^a\ \ ,
\label{lesser}
\end{equation}
where $\Gamma_{\beta}$ is the voltage dependent coupling parameter 
between probe $\beta$ and the scattering region\cite{wingreen,datta}. 
The self-consistent equations (\ref{gr}, \ref{poisson},\ref{lesser}) completely 
determine the nonlinear physics at the Hartree level.  We emphasize again that
the important departure of our theory from the familiar NEGF 
analysis\cite{wingreen,datta,stafford} is that we explicitly 
include the {\it internal} potential landscape $U({\bf r})$ into the 
Green's functions {\it self-consistently}. 

The net charge pile-up on a capacitor plate, measured from the 
equilibrium background, is derived from the right hand side of 
Eq. (\ref{poisson}). In the wideband limit\cite{meir} 
($\Gamma_\alpha=constant$)
\begin{equation}
\rho_\alpha (x)\ =\ \int \frac{dE}{2\pi} \Gamma_\alpha 
\left[ G^r G^a f(E-qV_\alpha) -G_o^r G_o^a f(E)\right]_{xx}\ \ ,
\label{rho2}
\end{equation}
where $G_o^{r,a}$ are the equilibrium Green's functions.
Next, we formally expand $G^r$ and $G^a$ in a power series of the internal
potential $U$, and expand the Fermi function in series of the bias voltage 
$V_\alpha$. Collecting terms according to the powers of $U$ and $V_\alpha$, 
Eq. (\ref{rho2}) reduces to the following infinite series which 
can be exactly summed,
\begin{eqnarray}
\rho_\alpha(x) &=& \left[\frac{d\sigma_\alpha}{dE}
(V_\alpha - U) + \frac{1}{2} \frac{d^2\sigma_\alpha}{dE^2} (V_\alpha-U)^2 +
\cdots\right]_{xx} \nonumber \\
&\equiv& \sigma_\alpha(E+V_\alpha-U) -\sigma_\alpha(E)
\label{eq18}
\end{eqnarray}
where the quantity $d\sigma_\alpha/dE$ is defined as
\begin{equation}
\frac{d\sigma_\alpha}{dE}\ \equiv\ -\int
\frac{dE}{2\pi}\left[G_o^r\Gamma_\alpha\frac{\partial f}{\partial E}
G_o^a\right]_{xx}\ \ .
\label{dnde1}
\end{equation}
The physical significance of the quantity $d\sigma_\alpha/dE$ can be
identified as the linear spatial dependent local partial density of 
states\cite{but1,ma} (LPDOS). Expression (\ref{dnde1}) has been obtained
before\cite{buttiker0,ma}.
For a conductor which is weakly coupled
to external leads, LPDOS gives the local DOS of this
conductor. Hence the spectral function $\sigma_\alpha(E)$ 
characterizes the local electronic structure of a nano-scale conductor.

The nonlinear charge distribution gives the general 
electrochemical capacitance versus voltage curve,
\begin{equation}
C = [\sigma_1(E+V_1-U_1)-\sigma_1(E)]/(V_1-V_2)
\label{cv}
\end{equation}
where we have set electron charge $q$ to be unity. To determine C we must
obtain internal potentials $U_1$ and $U_2$ at the two plates. For
this purpose we introduce the {\it geometrical} capacitance\cite{but6} 
$C_0$
\begin{equation}
C_0 = [\sigma_1(E+V_1-U_1)-\sigma_1(E)]/(U_1-U_2)
\label{U1}
\end{equation}
and
\begin{equation}
C_0 = -[\sigma_2(E+V_2-U_2)-\sigma_2(E)]/(U_1-U_2)
\label{U2}
\end{equation}
For a parallel plate capacitor, $C_0=A/(4\pi a)$ where $A$ is the area
of the plates and $a$ is their separation. In general $C_0$ can be
calculated numerically. The two equations (\ref{U1},\ref{U2})
determine the internal potentials $U_1$ and $U_2$ when the scattering 
spectral function $\sigma_\alpha$ and $C_0$ are known. 
Eqs. (\ref{cv}, \ref{U1}, \ref{U2}) gives, for the first time, 
the general electrochemical C-V curve for quantum capacitors. It is universal 
in the sense that system specific parameters only appear in the scattering 
spectral functions of the conductors. From these general results, two 
useful applications follow: 

\noindent
{\underline{\bf C-V curve}}. Our general results allow us to predict
capacitance-voltage curves.  For this application we note that
there are well established
methods for calculating the scattering LPDOS\cite{but2,ma}: 
either by Eq. (\ref{dnde1}) after evaluating the Green's
functions, or using the scattering wavefunctions\cite{but2} $\psi$ 
$d\sigma_\alpha/dE=|\psi|^2/{hv}$, where $v \sim \sqrt{E}$ is the velocity
and $h$ the Planck constant. Hence by solving a quantum scattering problem 
one obtains $\sigma_\alpha$. Let's consider a case where $\psi$ is not 
very sensitive to $E$, thus $d\sigma_\alpha(E)/dE\approx b_\alpha/ (2\sqrt{E})$ 
or $\sigma_\alpha(E)=b_\alpha\sqrt{E}$ where $b_\alpha$ is a constant. 
For this LPDOS, solving Eqs. (\ref{cv}, \ref{U1}, \ref{U2}) we obtain
\begin{equation}
C(V_1-V_2)=\frac{\beta-\sqrt{\beta^2-4(b_1^2-b_2^2) b_1^2 b_2^2
(V_1-V_2)}}{2(b_1^2-b_2^2)(V_1-V_2)}
\label{cv1}
\end{equation}
where
\begin{equation}
\beta=4\pi a b_1^2 b_2^2 +2b_1 b_2 (b_1+b_2) \sqrt{E}\ \ .
\end{equation}
The inset of Fig. (1) shows this electrochemical C-V curve as a function of 
$V\equiv (V_2-V_1)$ for two sets of parameters $b_\alpha$. The physical
reason that $C$ changes with $V$ is because the plates have finite DOS. Indeed, by 
making DOS very large ($b_1,b_2 \rightarrow \infty$) the voltage dependence of 
(\ref{cv1}) disappears and $C$ becomes purely geometrical.  Furthermore,
it can be confirmed that formula (\ref{cv1}) recovers the linear\cite{but1}
and second order nonlinear\cite{ma} capacitance coefficients when we take the
$(V_2-V_1) \rightarrow 0$ limit.

\noindent
{\underline{\bf Quantum scanning capacitance microscopy}}. 
Our general results suggest a quantum capacitance microscopy (QSCM).
This idea naturally follows from the results presented above: 
since the electrochemical capacitance varies with bias due to 
a finite DOS of the conductors involved, we should be able to find the 
DOS by measuring $C$. Essentially we wish 
to obtain spectral function $\sigma_2(E)$ or local density of states 
$d\sigma_2/dE$ as a function of energy for an unknown conductor, from a 
known $\sigma_1(E)$ of our QSCM ``tip'' which has been calibrated\cite{foot}. 
As the QSCM tip is scanned along the surface of a nano-scale conductor, 
or along the surface of a planar dielectric layer with nano-conductors 
buried underneath, experimentally one can measure the $C(V)$ curves at 
each spatial position.

From Eq.(\ref{cv}), we obtain $U_1$ as a function of potential difference 
$V$ using the known $\sigma_1$ and the measured $C(V)$, by solving the 
equation $\sigma_1(1-U_1)-\sigma_1(1)=VC(V)$, where we have set $V_1=0$, 
$V_2=V$, and $E=1$ is set at the Fermi energy of the QSCM ``tip'' 
($C(V)$ is measured at Fermi energy of the ``tip'' $E_F^1$, 
{\it e.g.} Eq. (\ref{cv1})).
Next, From Eq.(\ref{U1}) we obtain $U_2(V)$. With $U_1$ and $U_2$
we finally find $\sigma_2(E)$ from Eq. (\ref{U2}). In particular we can solve 
the spectral function $\sigma_2$ by representing it into a polynomial: 
\begin{equation}
\sigma_2(X) = \sum_{m=0}^n y_m X^m\ \ ,
\label{sigma2}
\end{equation}
where the coefficients $y_m \equiv (d^m\sigma_2/dX^m)/(m!)$ are 
just the linear ($m=1$) LPDOS and nonlinear ($m>1$) LPDOS\cite{ma}.
They are obtained by solving the following set of linear algebraic 
equations which come from Eq. (\ref{U2}),
\begin{equation}
U_2(V_j)-U_1(V_j) = 4\pi a \sum_{m=0}^n y_m\left[(1+V_j-U_2(V_j))^m -1\right]
\label{yi}
\end{equation}
where $j=1,2, \cdots,n$. Hence by making experimental measurements at 
$n$ different voltages $V_j$, we obtain the functional form of
$\sigma_2(E)$ from (\ref{sigma2}). 
Fig. (1) demonstrates the principle of QSCM. We use 
Eq. (\ref{cv1}) as the experimentally measured C-V curve (the inset) to
simulate a measurement. Then using the QSCM ``measured'' $\sigma_2(E)$ 
from Eq. (\ref{sigma2}), we plot the local density of states $d\sigma_2/dE$
versus energy $E$. The solid line is the exact $d\sigma_2/dE=b_2/(2\sqrt{E})$
and the dots are the QSCM result. We used 10 voltages in solving
Eq. (\ref{yi}) and the outcome is quite good, while using 3 voltages it
already represents a rough trend. 

In summary, we have developed a general nonlinear DC theory which is applied 
to investigate, for the first time, the full nonlinear charge distribution 
in nano-scale conductors. We have derived an exact expression of the 
electrochemical capacitance versus external bias voltage curve for 
quantum capacitors. This result is generic in the sense that all system 
specific information are included in the scattering 
local density of states. Hence the C-V formula has a wide range of 
applicability. By inverting this formula, we propose a novel QSCM. The 
QSCM extends the ability of the usual scanning capacitance 
microscopy\cite{foot2}: QSCM includes the quantum corrections to capacitance 
in mapping out the the spatial charge distribution; and it gives the local 
density of states as a function of electron energy. We believe such an 
idea can be readily implemented in a scanning apparatus using tiny tips as the 
calibrated conductor, thus allowing measurements of electronic 
properties for other conductors at the nano-scale.

\noindent
{\bf Acknowledgments.}
We gratefully acknowledge support by a RGC grant from the SAR Government 
of Hong Kong under grant number HKU 7112/97P, and a CRCG grant from the 
University of Hong Kong. H. G is supported by NSERC of Canada and FCAR 
of Qu\'ebec.

\section*{Figure Caption}

\begin{itemize}

\item[{Fig.(1)}]
Operation of QSCM: comparison of ``measured'' LPDOS to the exact one.  
Solid lines: exact; solid circles: fitted LPDOS using 10 voltages; solid
triangles: fitted LPDOS using 3 voltages. Upper set of curves are for 
$b_2=0.002$, lower set $b_2=0.001$.  The QSCM tip has been fixed with
$b_1=0.003$.  The energy unit is the Fermi energy $E_F^1$ of the tip.
Inset: the electrochemical capacitance versus voltage curve for the two
sets of $b_2$: upper curve is for $b_2=0.002$.  The unit of $V$ is
$E_F^1/e$.
\end{itemize}

\begin{thebibliography}{00}
\bibitem{foot}
The calibration can be done in many ways. For example one can 
prepare two identical conductors thus $\sigma_1(E)=\sigma_2(E)$, and
solve for $\sigma_1$.
\end{thebibliography}
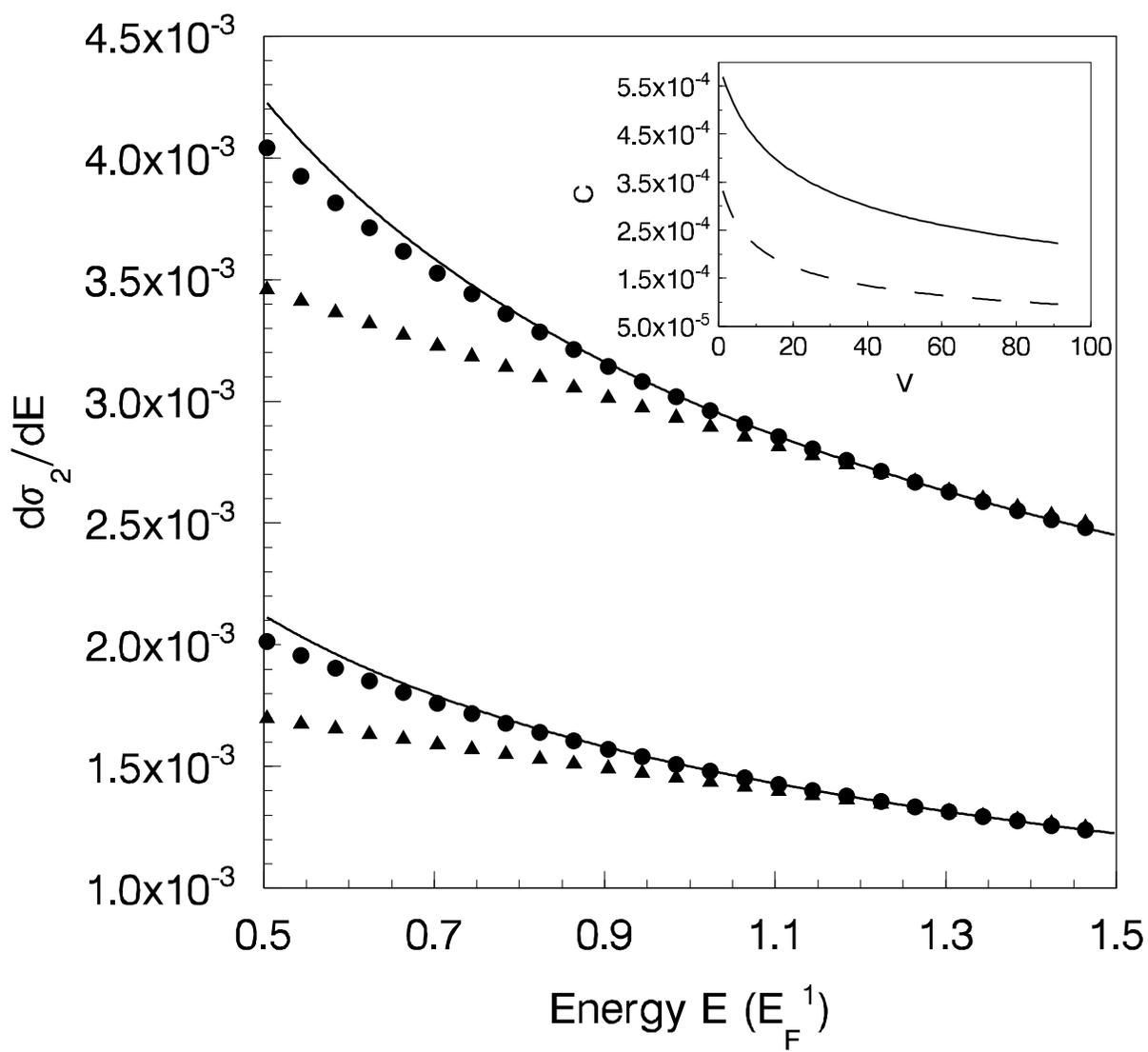
\end{document}